\pdfoutput=1
\documentclass[a4paper,twoside]{article}
\usepackage{balance}
\usepackage{epsfig}
\usepackage{subcaption}
\usepackage{calc}
\usepackage{amssymb}
\usepackage{amstext}
\usepackage{amsmath}
\usepackage{amsthm}
\usepackage{multicol}
\usepackage{pslatex}
\usepackage{apalike}
\usepackage{algorithm2e}
\usepackage[bottom]{footmisc}
\usepackage[acronym,shortcuts]{glossaries}
\usepackage{multirow}
\usepackage{SCITEPRESS} 
\newacronym{AdaBoost}{AdaBoost}{adaptive boosting}
\newacronym{BERT}{BERT}{bidirectional transformer}
\newacronym{BOW}{BOW}{bag Of words}
\newacronym{CNN}{CNN}{convolutional neural network}
\newacronym{DT}{DT}{decision tree}
\newacronym{GPT-2}{GPT-2}{generative pre-trained transformer 2}
\newacronym{HTML}{HTML}{hyper text markup language}
\newacronym{KNN}{KNN}{k-nearest neighbor}
\newacronym{LR}{LR}{logistic regression}
\newacronym{NB}{NB}{na\"ive Bayes}
\newacronym{OCR}{OCR}{optical character recognition}
\newacronym{RF}{RF}{random forest}
\newacronym{SVM}{SVM}{support vector machine}
\newacronym{SMTP}{SMTP}{simple mail transfer protocol}
\newacronym{TF-IDF}{TF-IDF}{term frequency-inverse document frequency}
\newacronym{VBSF}{VBSF}{visual based spam filter}
\newacronym{XGBoost}{XGBoost}{extreme gradient boosting}
\newacronym{XLNet}{XLNet}{extreme language understanding network}

\begin{document}
\title{VBSF: A Visual-Based Spam Filtering Technique for Obfuscated Emails}

\author{\authorname{Ali Hossary\sup{1} \orcidAuthor{0009-0002-9227-2662} , Stefano Tomasin\sup{1}\orcidAuthor{0000-0003-3253-6793}}
\affiliation{\sup{1}Dept. of Information Engineering (DEI), University of Padova, Italy}
\email{\{ali.hossary, stefano.tomasin\}@unipd.it}
}

\keywords{Spam Email Detection, Hidden Text Salting, Obfuscated Words, Email Rendering, Machine Learning, Optical Character Recognition (OCR), Convolutional Neural Networks (CNN), Ensemble Learning.}

\abstract{Recent spam email techniques exploit visual effects in text messages, such as poisoning text, obfuscating words, and hidden text salting techniques. These effects were able to evade spam detection techniques based on the text. In this paper, we overcome this limitation by introducing a novel visual-based spam detection architecture, denoted as VBSF. The multi-step process mimics the human eye's natural way of processing visual information, automatically rendering incoming emails and capturing their content as it appears on a user screen. Then, two different processing pipelines are applied in parallel. The first pipeline pertains to the perceived textual content, as it includes OCR to extract rendered textual content, followed by NB and DT content classifiers. The second pipeline focuses on the appearance of the email, as it analyzes and classifies the images of rendered emails through a specific convolutional neural network. Lastly, a meta classifier integrates text- and image-based classifier outputs exploiting the stacking ensemble learning method. The performance of the proposed VBSF is assessed, showing that It achieves an accuracy of more than 98\%, which is higher than the compared existing techniques on the designed dataset.}

\onecolumn \maketitle \normalsize \setcounter{footnote}{0} \vfill

\section{\uppercase{Introduction}}
\label{sec:introduction}

Emails have witnessed an overwhelming global volume exceeding 200 billion messages daily. However, a staggering 80-90\% of this flow comprises spam, which both annoys users and fuels malicious activities like phishing, fraud, and malware dissemination. Conventional anti-spam methods like blacklists and heuristics struggle against this onslaught, prompting the development of scalable and adaptive techniques. Machine learning is the leading approach, with state-of-the-art classifiers achieving over 90\% accuracy. The prominence of machine learning in spam defense is driven by factors such as the massive scale of global spam, evolving spam tactics, and ongoing algorithmic advancements tailored for text analysis. Moreover, the exponential growth in computational resources enhances the effectiveness of spam filtering models. 
 
Among advanced spam techniques, visual effects on text messages are particularly challenging for spam detectors, since they defeat text-based detection systems. Such approaches go under the names poisoning text, obfuscated words, and {\em hidden salting} \cite{bergholz2008detecting}. Recently, \cite{Sokolov_Olufowobi_Herndon_2020a} has shown that spam detection techniques can be evaded by replacing some characters with others that look very similar but come from a different alphabet. For a review of hidden salting tricks see \cite{janez2023review}. 

This paper proposes new techniques to detect spam messages, including hidden salting and other visual attack strategies. The introduced solution is a \ac{VBSF} and it emulates the human visual perception of emails, thus aiming at reading the text as perceived by the human reader. Therefore, visual tricks such as fainted colors and small text used to hide part of the text and letting the reader see hidden spam content will also affect our detection technique that will be able to {\em see} the spam content and predict it exploiting multiple diverse classifiers. 

The findings underscore the importance of model composition and the value of incorporating diverse classifiers to achieve superior results. Our enhanced VBSF represents a promising advancement in predictive modeling, offering a pathway for further refinement and optimization of our approach. The resulting decision-making mechanism is robust and provides a final classification accuracy of the meta-classifier surpassing the accuracy of all the base models, exceeding 98\%.

The rest of this paper is organized as follows. In Section~II we review the existing literature with a focus on works related to our proposed VBSF solution. Section~III presents the VBSF technique in detail. We then design its implementation in Section~III. The performance results of the proposed solution and existing approaches are presented and discussed in Section~V. Lastly, Section~VI provides the main conclusions of this work.

\begin{figure*}
    \centering
   \includegraphics[width=0.9\linewidth]{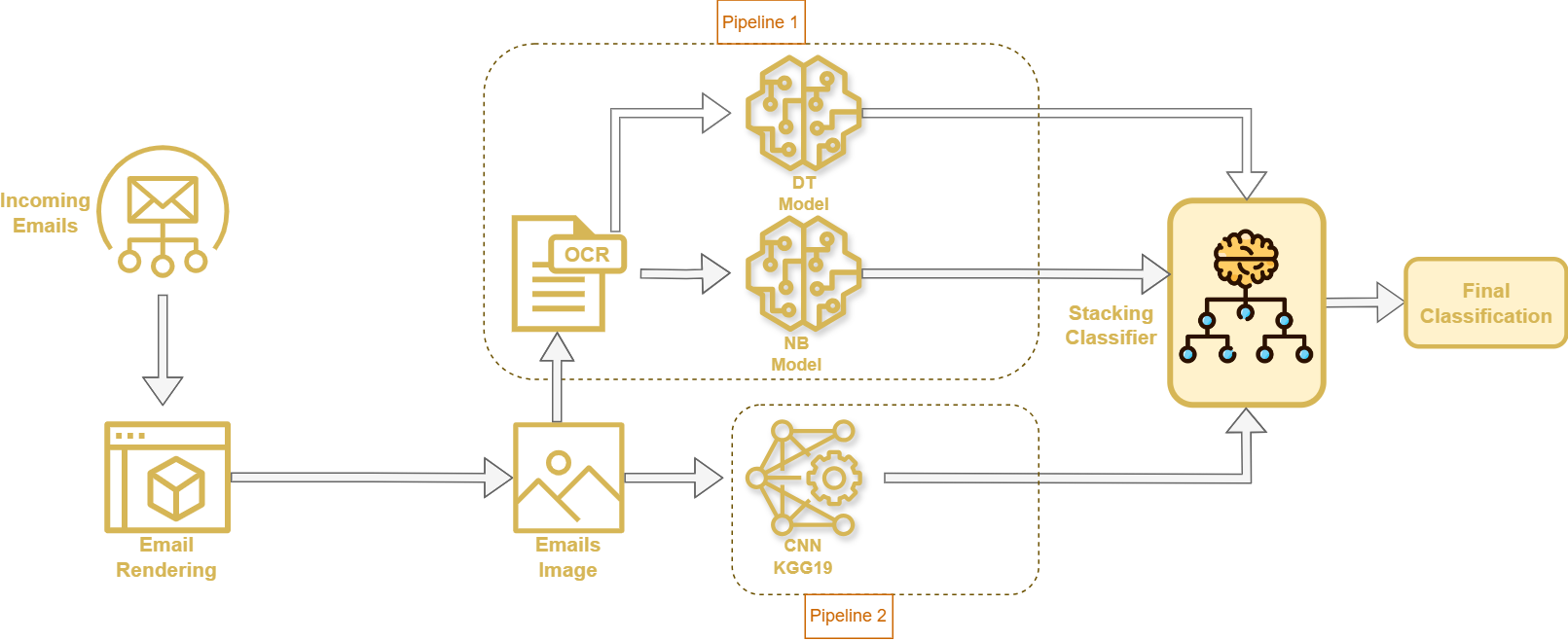}
    \caption{The proposed \ac{VBSF} solution.}
    \label{fig:pipeline}  
\end{figure*}

\section{\uppercase{Related Work}}

Several works have addressed the problem of detecting spam when hidden text salting is used. In many cases, machine learning techniques are employed, as they are well known to be effective in several security-related applications~\cite{9277523}.

In \cite{moens2010identifying}, the rendering process is tapped into. The rendering commands are analyzed to identify sections of the source text (plaintext) that will be invisible to human readers, based on criteria such as text character and background colors, font size, and overlapping characters. Furthermore, the visible text (cover text) is reconstructed from rendering commands, and the character reading order is identified, which may differ from the rendering order. In our study, we also render emails as images to analyze how they are perceived by the human eye. Instead of exploiting rendering commands, we render the whole content and then operate on the rendered image, thus being more flexible and independent of the specific rendering process. Moreover, we exploit powerful tools such as OCR and image analysis by neural networks that have proven to be effective in detection processes.

In \cite{electronics11132053} it has been proposed to use three sub-models to extract three features from images. In particular, two sub-models for text processing extract topic-based features (to identify the main subject of the message) and word-embedding-based features (to capture the meaning and relationships between words in the message) using the text contained in the images extracted by OCR. Then, a convolution-based sub-model extracts convolution-based features from images. Lastly, text and image features extracted from each sub-model are input into the classifier model that decides on the spam nature of the email. Thus, \cite{electronics11132053} proposes a technique for classifying spam images using image and text features extracted from images, which is related to our approach. We use instead three classifiers, each with its feature extraction method, followed by a stacking meta classifier to consolidate the predictions of sub-models, additionally. Moreover, the focus of \cite{electronics11132053} was on spam images included in emails, while we generated the images from the incoming emails.

In \cite{biggio2007image} the focus is on detecting spam techniques that hide the real content of the image. The proposed approach aims at identifying a specific characteristic of spam images with embedded text - the presence of content obscuring techniques. The underlying rationale is that images containing embedded text, which are deliberately obscured to render \ac{OCR} ineffective, are likely to be spam.

In \cite{naiemi2019efficient}, a method based on the histogram of oriented gradients, HOG, and a \ac{SVM} has been used for \ac{OCR} in images contained in emails. One of the limitations faced by the HOG feature extraction method is its lack of resistance against character variations on scales and translations.
 The proposed enhanced HOG feature extraction method has been used so that the \ac{OCR} system of spam has been enhanced by using the HOG feature extraction method in such a way to be both resistant against the character variations on scale and translation and to be computationally cost-effective. Our work focuses on text emails with hidden salting tricks, rather than on emails containing spam images.

Other approaches for spam detection using machine learning approaches include a bio-inspired technique \cite{9222163}, such as particle swarm optimizations and genetic algorithms which are used to optimize the performance of classifiers: it turned out that multinomial \ac{NB} with the genetic algorithm outperforms the other. Still, no visual tricks were considered in \cite{9222163}.

In \cite{9551992}, an unsupervised framework for spam detection is proposed, that resorts to a clustering approach including multiple algorithms. A suitable feature reduction is applied to obtain seven features that represent impactful analytical email characteristics from a multiangular point of view. However, this solution primarily uses the email content (body) and the subject header and does not properly deal with visual tricks. 

In our work, we also apply a \ac{CNN} directly to an image rendering an email. Such an approach has been adopted in other contexts (not related to spam detection). For example, in \cite{Rizky_Yudistira_Santoso_2023} a \ac{CNN} is used to recognize text in images. Several modifications of the images have been investigated and the best model turned out to be the VGG-16 architecture along with specific image transformations. The model architecture used in this study could be a valuable resource for developing future text detection systems.

\section{\uppercase{Visual-Based Filtering Technique}}
\label{sec:methodology}

Spammers often use \ac{HTML} techniques such as hidden or invisible text, \ac{HTML} comments, or misleading formatting to trick traditional text-based classifiers. At the same time, to convey the spam message, such tricks should provide a final image of the email that is clearly readable to the human reader. The basic idea of \ac{VBSF} is to first render the email as an image, then perform spam detection on the image: this enables the spam detector to operate on the same input as provided to the human reader. This approach enhances the classifier's ability to detect spam accurately, as it considers the visual presentation of the email, uncovering potential malicious elements that might be hidden in the \ac{HTML} code.

In detail, first, the email is rendered as an image: this includes interpreting HTML commands (on fonts, colors, and page layout), adding attached images, etc. Then, two spam detection techniques are applied in parallel on the obtained image, each denoted as a {\em pipeline}. The first pipeline is based on the {\em perceived content}, obtained through an \ac{OCR}: the extracted text content is then fed to a content-based classification system based on \ac{NB} and \ac{DT} classifiers. The latter pipeline is based on the {\em visual appearance} and classifies images of the email content by using a\ac{CNN}. 

Lastly, a meta classifier combines the outputs of both text- and image-based classifiers by a stacking ensemble learning method. Through experiments, we observed a remarkable increase in testing accuracy. Integrating the \ac{DT} classifier proved to be particularly impactful, contributing to a significant enhancement in predictive performance.

Fig.~\ref{fig:pipeline} shows the workflow of the proposed \ac{VBSF}, which is composed of the following elements:
\begin{itemize}
    \item Rendering of the email as an image
    \item First pipeline: text extraction using OCR, followed by content-based filters
    \item Second pipeline: image-based classifier (by a CNN model), applied to the email image
    \item Meta classifier, utilizing stacking ensemble method, as an ensemble learning technique
\end{itemize} 
Each element of the VBSF solution is described in detail below.

\subsection{Email Rendering}

The email rendering step generates the image of the email, as it would be shown to the end human reader. This includes the formatting of text and page according to the \ac{HTML} format, the inclusion of images, etc. Hidden salting tricks are also exploiting such formatting parts so that once the email is shown to the reader, it shows content (such as sentences or images) that are hard to identify in the original \ac{HTML} document~\cite{bergholz2008detecting}. The rendering can be easily obtained with one of the several tools that allow the conversion of an \ac{HTML} file into an image (as it was rendered in a browser or email reader). The obtained image is then processed in the forthcoming steps.

\subsection{First Pipeline: OCR Engine and Textual-Based Classificarion} The first pipeline aims at detecting spam by first converting the image of the email with \ac{OCR} and then applying spam detectors on the obtained text. Noting that this step is not the inverse of email rendering, as the text captured by the \ac{OCR} is very close to what the human eyes perceive, and can be very different from the textual content of the \ac{HTML} files, due to the hidden salting tricks.

\subsubsection{NB and DT Classifiers for Spam Filtering} Once the image has been converted into the {\em perceived text}, we apply text-based spam classifiers to detect the presence of spam. In particular, we consider the NB classifier as one of the most renowned and effective content-based spam filters. We also use a \ac{DT} classifier which operates in a different manner and is also very effective.

\subsection{Second Pipeline: Image-Based Classification With CNN}

The second pipeline aims to detect spam directly from the {\em appearance} of the email as rendered in the image, including colors and other visual objects. Here we use a specific computer vision CNN model as a very effective tool for image classification in similar contexts.
The CNN, serves as a visual perception model, learns patterns and features crucial for distinguishing between 
 the images of spam and ham emails

\begin{figure}[!h]
  \centering
   \includegraphics[width=7.5cm]{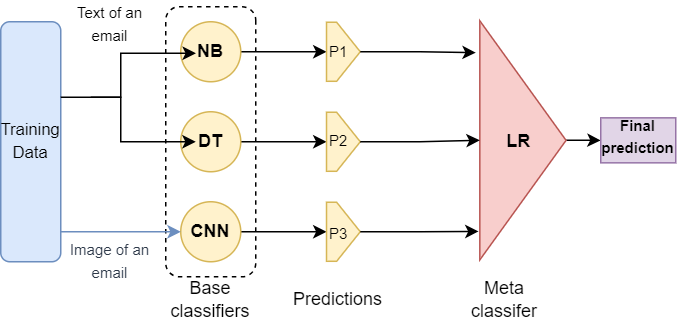}
  \caption{Stacking classifier architecture.}
  \label{fig:Meta Classifier}
 \end{figure}

 \subsection{Final Classification}

Since we have two pipelines aiming at providing a classification of an email, and to consolidate the classification predictions from both pipelines, which are predictions of baseline models, we incorporated a meta-classifier \cite{wolpert1992stacked} utilizing the stacking ensemble method, as depicted in Fig.~\ref{fig:Meta Classifier}.
The stacking classifier architecture elements are as follows:
\begin{itemize}
    \item Baseline classification models: the NB, DT and CNN models.
    \item Predictions of the base models: the training predictions of the models in binary form, also called stacking features, which are fed as input training data for the meta classifier
    \item Meta classifier: a \ac{LR} classifier is chosen, trained on the training predictions of the base classifiers, in its output the final binary prediction of the whole architecture.
\end{itemize}

The \ac{LR} classifier is fed with the predictions generated by the \ac{CNN} classifier, \ac{DT}classifier, and the \ac{NB} classifier, the use of a diverse set of base classifiers, leveraged their diverse nature to improve final predictive accuracy.

The utilization of Logistic Regression classifier as the stacking classifier further refines the integration process, providing a well-balanced synthesis of the predictions from baseline models, ensuring a more comprehensive and nuanced analysis of the input data, leading to a more reliable and informed final classification outcome. Interestingly, we experimented with various models as potential stacking meta-classifiers, such as \ac{SVM}, \ac{DT}, and random forests. However, after rigorous evaluation, \ac{LR} emerged as the best fit.

\section{\uppercase{VBSF Environment Setup}}
\label{sec:experiments}

In this section, we describe our dataset and the Environment of the \ac{VBSF} technique including fine tuning the Neural Network model process and its setup used for performance evaluation in the next section.

\subsection{Dataset Collection And Preparation}

A mix of publicly available datasets has been used together with a dataset of hand-crafted emails for specific testing purposes. In particular, we considered the Enron 1 and Enron 4 and pre-processed Spam Assasin email corpus. The combination of parts of the three datasets was the best fit for our proposed model and led to better generalization: indeed, Enron 1 and 4 have enough textual features, while they lack colors and visual features, while Spam Assasin is rich in colors and visual emails but alone was not big enough. 

The combined dataset had imbalanced class distributions (40\% spam and 60\% ham emails), so we increased the number of spam emails to balance the dataset and prevent overfitting toward the majority class. As a result, we have 4009 benign emails (ham) and 3800 spam emails. Additionally, a few samples have been minimally modified by applying some spam tricks, trying to emulate an adversary behavior, such as spam word spacing using \ac{HTML} comments, ham and spam word injection, and modification of the size and bold effects. Most of these samples succeeded in misleading existing classifiers.

\begin{table*}[h] 
\caption{Accuracy of VBSF-Pipeline 1 (after applying OCR) vs existing normal Text-Based Detection (without OCR)}
\label{tab:accuracy_comparison}
\centering
\begin{tabular}{|c|c|c|}
\hline
\multirow{2}{*}{\textbf{ML Model}} & \multicolumn{2}{c|}{\textbf{Accuracy}} \\ \cline{2-3}
& \textbf{Text-based spam detector} & \textbf{VBSF - Pipeline 1} \\ \hline
\textbf{NB} & 94 \% & 96 \% \\ \hline
\textbf{DT} & 95 \% & 97 \% \\ \hline
\textbf{LR} & 96 \% & 97 \% \\ \hline
\textbf{SVM} & 80 \% & 96 \% \\ \hline
\textbf{AdaBoost} & 96 \% & 96 \% \\ \hline
\textbf{KNN} & 89 \% & 91 \% \\ \hline
\end{tabular}
\end{table*}

\subsection{VBSF Setup}

For the OCR of the obtained image, we resort to Google Tesseract \cite{Tesseract-Ocr} \cite{PyPI}.  

\begin{figure}[!h]   
\centering
   \includegraphics[width=7.8cm]{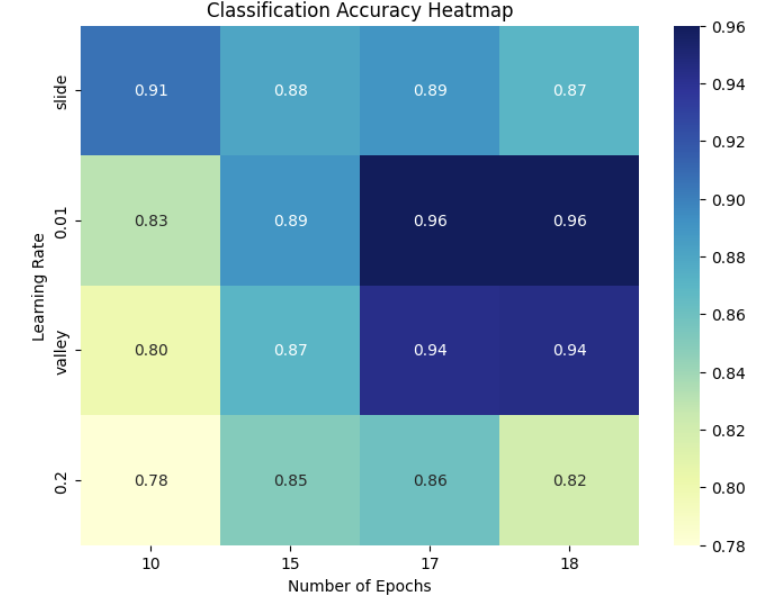}
\caption{Classification accuracy heatmap for the VGG19 model.}     \label{fig:Heat map}   \end{figure}

For the second pipeline that classifies the image into the two spam and ham classes, we resort to the 
CNN VGG-19 neural network that utilizes small $3 \times 3$ filters across all convolutional layers, resulting in optimal performance reflected in its low error rate \cite{Zheng_Yang_Merkulov_2018}.
We fine-tuned the VGG-19 model through various settings to attain optimal predictive performance. Several key hyperparameters were precisely adjusted, including the learning rate, the number of training and fine-tuning epochs. Additionally, we incorporated data augmentation techniques to further enhance the model's ability to generalize patterns from our dataset.
Choosing an appropriate learning rate was a critical step in achieving the best results. Fig.~\ref{fig:Heat map} shows the classification accuracy heatmap for the VGG19 model. We note that through a systematic exploration of various learning rates and epoch numbers, we identified the spot that led to a remarkable accuracy and low validation loss, without overfitting or underfitting the data.

\section{\uppercase{Numerical Results}}

We first assess the accuracy of the first pipeline {\em without the DT branch} and with different classifiers on the OCR output. In particular, we consider the \ac{NB}, \ac{DT}, \ac{LR}, \ac{SVM}, AdaBoost \cite{FREUND1997119}, and \ac{KNN} classifiers. For comparison purposes, we also apply the same classifiers directly to the original \ac{HTML} text file, thus without passing through the visual representation and OCR. 

Table~\ref{tab:accuracy_comparison} shows the accuracy of both the VBSF and text-based detectors for the various classifiers. Interestingly, when applied to the raw emails (source etext) in our new dataset, the performance of conventional NB and DT classifiers did not match those on well-known email datasets. However, when we used the NB and DT classifiers on emails that went through the OCR after rendering and capturing, both classifiers demonstrated a remarkable performance improvement. The adaptation of OCR technology appeared to enhance the classifier's ability to discern spam characteristics within the text, showcasing the versatility of the NB and DT classifiers in the context of our VBSF's first pipeline. This nuanced observation underscores the importance of tailoring spam filters to the unique characteristics of the dataset at hand, optimizing their performance for diverse sources and formats of email content.

\begin{table*}[h]
    \caption{VBSF Accuracy With Different Meta Classifiers}
    \label{tab:stacking-classifiers-3}
    \centering
    \begin{tabular}{|c|c|c|c|}
        \hline
        \textbf{Meta Classifier Model} & \textbf{VBSF final} & \textbf{False Positive}  & \textbf{False Negative} \\
        \textbf{used for VBSF} & \textbf{test accuracy} & \textbf{Rate} & \textbf{Rate} \\
        \hline
        LR & 98.3\% & 1.2\%  & 0.5\%\\
        \hline
        Random Forest & 97.3\% & 1.7\%  & 1.0\% \\
        \hline
        DT & 96.6\% & 2.3\%  & 1.1\% \\
        \hline
    \end{tabular}
\end{table*}

Now, we assess the performance of the VBSF solution. Table~\ref{tab:stacking-classifiers-3} shows the test accuracy of the meta classifier after augmenting the first pipeline of the VBSF. Several meta-classifiers underwent testing, again, among which \ac{LR} produced superior performance compared to others reaching more than 98\% accuracy, hence it was selected as the preferred choice.

Through experimentation and evaluation, we observed a remarkable increase in testing accuracy. The integration of the \ac{DT} classifier proved to be particularly impactful, contributing to a significant enhancement in predictive performance. These findings underscore the importance of model composition and the value of incorporating diverse classifiers to achieve superior results. Our enhanced variant of VBSF represents a promising advancement in predictive modeling, offering a pathway for further refinement and optimization of our approach.

\section{\uppercase{Conclusions}}
\label{sec:conclusions}

We have proposed a new approach to detect emails that use visual tricks (or hidden salting tricks)and HTML-related tricks, to convey spam messages to end users. By employing a multi-step process imitating the natural processing of visual information by the human eye, alongside text extraction of email snapshots using OCR followed by textual content classification using an NB classifier, augmented by a \ac{DT} classifier, our system efficiently cleans and analyzes email text content. Moreover, integrating a CNN as a visual perception classification model enhances the system's ability to discern between spam and legitimate emails based on visual features and cues.

A remarkable strength of our proposed solution lies in its adaptability to the dynamic nature of spamming techniques, especially the visual ones. The proposed model includes parsing all HTML tags and formatting the content according to their specifications. Whether it's normal content, known spam content hiding tricks, or crafty spam tactics, all elements are visually visible and ready for further investigation.  
By integrating text-based and image-based classifiers in a meta-classifier using stacking ensemble learning, our system achieves a very good final classification accuracy exceeding 98\%. This holistic approach enhances both the accuracy and the resilience against evolving spam tactics.

\bibliographystyle{apalike}
{\small
\bibliography{PaperBibliography}}

\end{document}